\documentclass[a4paper,11pt]{article}
\usepackage[mathlines]{lineno}

\usepackage{jcappub}

\usepackage[table,svgnames,dvipsnames]{xcolor}
\usepackage[normalem]{ulem}
\usepackage{aas_macros}
\usepackage{hyperref}
\usepackage{verbatim}
\usepackage{tabularx}
\usepackage{orcidlink} % for author list
\usepackage{hanging} % for affiliations 
\usepackage{multirow}
\usepackage{makecell}
\usepackage{siunitx}
\usepackage{dcolumn} % Align table columns on decimal point
\usepackage{bm} % bold math
\usepackage{xspace}
\usepackage{graphicx} % Include figure files
\usepackage{subfigure}
\usepackage{cases} % for dealing with mathematics,
\usepackage{booktabs}

\usepackage[nameinlink,noabbrev]{cleveref}
\crefname{equation}{Eq.}{Eqs.}
\crefname{section}{Section}{Sections}
\crefname{figure}{Figure}{Figures}
\crefname{table}{Table}{Tables}
\crefname{appendix}{Appendix}{Appendices}
\Crefname{figure}{Figure}{Figures}
\Crefname{equation}{Equation}{Equations}
\Crefname{section}{Section}{Sections}
\Crefname{table}{Table}{Tables}

\newcommand{\hl}[1]{#1}

\newcommand{\mpchcubed}{h^{-3}\,\mathrm{Mpc}^3}
\newcommand{\hmpc}{h\,\mathrm{Mpc}^{-1}}

\def\2pr{^{\prime \prime}}

\def\geqsim{\lower.73ex\hbox{$\sim$}\llap{\raise.4ex\hbox{$>$}}$\,$}
\def\leqsim{\lower.73ex\hbox{$\sim$}\llap{\raise.4ex\hbox{$<$}}$\,$}
\newcommand{\fnl}{f^{\rm loc}_{\rm NL}}

\newcommand{\sysnet}{\textsc{SYSNet} }

\newcommand{\minuit}{\textsc{minuit}}
\newcommand{\desilike}{\textsc{desilike}}
\newcommand{\regressis}{\textsc{Regressis} }

\newcommand{\elgxlrg}{\rm ELGxLRG}
\newcommand{\elgxqso}{\rm ELGxQSO}
\newcommand{\lrgxqso}{\rm LRGxQSO}
\newcommand{\exl}{\rm ExL}
\newcommand{\exq}{\rm ExQ}
\newcommand{\lxq}{\rm LxQ}

\newcommand{\zrqso}{0.8<z<3.1}
\newcommand{\zrelgxlrg}{0.8<z<1.1}
\newcommand{\zrelgxqso}{0.8<z<1.6}
\newcommand{\zrlrgxqso}{0.8<z<1.1}
\newcommand{\zeff}{z_{\rm eff}}
\newcommand{\wfkp}{w_{\rm FKP}}

\title{Local primordial non-gaussianity using cross-correlations of DESI tracers}

\author[a]{{A.~J.~Rosado-Mar\'{i}n}\orcidlink{0000-0001-7545-3504},}
\author[b]{{A.~J.~Ross}\orcidlink{0000-0002-7522-9083},}
\author[a]{{H.~Seo}\orcidlink{0000-0002-6588-3508},}
\author[c]{{E.~Chaussidon}\orcidlink{0000-0001-8996-4874},}
\author[c]{{J.~Aguilar},}
\author[d]{{S.~Ahlen}\orcidlink{0000-0001-6098-7247},}
\author[e,f]{{D.~Bianchi}\orcidlink{0000-0001-9712-0006},}
\author[g]{{D.~Brooks},}
\author[c]{{T.~Claybaugh},}
\author[c]{{A.~Cuceu}\orcidlink{0000-0002-2169-0595},}
\author[h]{{A.~de la Macorra}\orcidlink{0000-0002-1769-1640},}
\author[i]{{A.~de Mattia}\orcidlink{0000-0003-0920-2947},}
\author[j]{{R.~Demina},}
\author[k,l]{{B.~Dey}\orcidlink{0000-0002-5665-7912},}
\author[g]{{P.~Doel},}
\author[c,m]{{S.~Ferraro}\orcidlink{0000-0003-4992-7854},}
\author[n,o]{{A.~Font-Ribera}\orcidlink{0000-0002-3033-7312},}
\author[p,q]{{J.~E.~Forero-Romero}\orcidlink{0000-0002-2890-3725},}
\author[r,s,t]{{E.~Gaztañaga}\orcidlink{0000-0001-9632-0815},}
\author[u]{{S.~{Gontcho A Gontcho}}\orcidlink{0000-0003-3142-233X},}
\author[v]{{G.~Gutierrez},}
\author[w]{{C.~Hahn}\orcidlink{0000-0003-1197-0902},}
\author[x,i]{{H.~K.~Herrera-Alcantar}\orcidlink{0000-0002-9136-9609},}
\author[y,z]{{D.~Huterer}\orcidlink{0000-0001-6558-0112},}
\author[aa]{{M.~Ishak}\orcidlink{0000-0002-6024-466X},}
\author[ab]{{R.~Joyce}\orcidlink{0000-0003-0201-5241},}
\author[ac]{{D.~Kirkby}\orcidlink{0000-0002-8828-5463},}
\author[c]{{A.~Kremin}\orcidlink{0000-0001-6356-7424},}
\author[g]{{O.~Lahav}\orcidlink{0000-0002-1134-9035},}
\author[ad]{{C.~Lamman}\orcidlink{0000-0002-6731-9329},}
\author[c]{{M.~Landriau}\orcidlink{0000-0003-1838-8528},}
\author[c]{{M.~E.~Levi}\orcidlink{0000-0003-1887-1018},}
\author[ae,o]{{M.~Manera}\orcidlink{0000-0003-4962-8934},}
\author[ab]{{A.~Meisner}\orcidlink{0000-0002-1125-7384},}
\author[n,o]{{R.~Miquel},}
\author[s]{{S.~Nadathur}\orcidlink{0000-0001-9070-3102},}
\author[l]{{J.~ A.~Newman}\orcidlink{0000-0001-8684-2222},}
\author[i,c]{{N.~Palanque-Delabrouille}\orcidlink{0000-0003-3188-784X},}
\author[af,ag,ah]{{W.~J.~Percival}\orcidlink{0000-0002-0644-5727},}
\author[ai]{{F.~Prada}\orcidlink{0000-0001-7145-8674},}
\author[aj]{{I.~P\'erez-R\`afols}\orcidlink{0000-0001-6979-0125},}
\author[ak]{{G.~Rossi},}
\author[al]{{E.~Sanchez}\orcidlink{0000-0002-9646-8198},}
\author[c]{{D.~Schlegel},}
\author[c]{{J.~Silber}\orcidlink{0000-0002-3461-0320},}
\author[z]{{G.~Tarl\'{e}}\orcidlink{0000-0003-1704-0781},}
\author[ab]{{B.~A.~Weaver},}
\author[am]{{C.~Zhao}\orcidlink{0000-0002-1991-7295},}
\author[an]{{H.~Zou}\orcidlink{0000-0002-6684-3997}}

\affiliation[a]{Department of Physics \& Astronomy, Ohio University, 139 University Terrace, Athens, OH 45701, USA}
\affiliation[b]{Department of Physics, The Ohio State University, 191 West 19 Woodruff Avenue, Columbus, OH 43210, USA}
\affiliation[c]{Lawrence Berkeley National Laboratory, 1 Cyclotron Road, Berkeley, CA 94720, USA}
\affiliation[d]{Department of Physics, Boston University, 590 Commonwealth Avenue, Boston, MA 02215 USA}
\affiliation[e]{Dipartimento di Fisica ``Aldo Pontremoli'', Universit\`a degli Studi di Milano, Via Celoria 16, I-20133 Milano, Italy}
\affiliation[f]{INAF-Osservatorio Astronomico di Brera, Via Brera 28, 20122 Milano, Italy}
\affiliation[g]{Department of Physics \& Astronomy, University College London, Gower Street, London, WC1E 6BT, UK}
\affiliation[h]{Instituto de F\'{\i}sica, Universidad Nacional Aut\'{o}noma de M\'{e}xico,  Circuito de la Investigaci\'{o}n Cient\'{\i}fica, Ciudad Universitaria, Cd. de M\'{e}xico  C.~P.~04510,  M\'{e}xico}
\affiliation[i]{IRFU, CEA, Universit\'{e} Paris-Saclay, F-91191 Gif-sur-Yvette, France}
\affiliation[j]{Department of Physics \& Astronomy, University of Rochester, 206 Bausch and Lomb Hall, P.O. Box 270171, Rochester, NY 14627-0171, USA}
\affiliation[k]{Department of Astronomy \& Astrophysics, University of Toronto, Toronto, ON M5S 3H4, Canada}
\affiliation[l]{Department of Physics \& Astronomy and Pittsburgh Particle Physics, Astrophysics, and Cosmology Center (PITT PACC), University of Pittsburgh, 3941 O'Hara Street, Pittsburgh, PA 15260, USA}
\affiliation[m]{University of California, Berkeley, 110 Sproul Hall \#5800 Berkeley, CA 94720, USA}
\affiliation[n]{Instituci\'{o} Catalana de Recerca i Estudis Avan\c{c}ats, Passeig de Llu\'{\i}s Companys, 23, 08010 Barcelona, Spain}
\affiliation[o]{Institut de F\'{i}sica d’Altes Energies (IFAE), The Barcelona Institute of Science and Technology, Edifici Cn, Campus UAB, 08193, Bellaterra (Barcelona), Spain}
\affiliation[p]{Departamento de F\'isica, Universidad de los Andes, Cra. 1 No. 18A-10, Edificio Ip, CP 111711, Bogot\'a, Colombia}
\affiliation[q]{Observatorio Astron\'omico, Universidad de los Andes, Cra. 1 No. 18A-10, Edificio H, CP 111711 Bogot\'a, Colombia}
\affiliation[r]{Institut d'Estudis Espacials de Catalunya (IEEC), c/ Esteve Terradas 1, Edifici RDIT, Campus PMT-UPC, 08860 Castelldefels, Spain}
\affiliation[s]{Institute of Cosmology and Gravitation, University of Portsmouth, Dennis Sciama Building, Portsmouth, PO1 3FX, UK}
\affiliation[t]{Institute of Space Sciences, ICE-CSIC, Campus UAB, Carrer de Can Magrans s/n, 08913 Bellaterra, Barcelona, Spain}
\affiliation[u]{University of Virginia, Department of Astronomy, Charlottesville, VA 22904, USA}
\affiliation[v]{Fermi National Accelerator Laboratory, PO Box 500, Batavia, IL 60510, USA}
\affiliation[w]{Department of Astronomy, University of Texas at Austin, 2515 Speedway, TX 78712, USA}
\affiliation[x]{Institut d'Astrophysique de Paris. 98 bis boulevard Arago. 75014 Paris, France}
\affiliation[y]{Department of Physics, University of Michigan, 450 Church Street, Ann Arbor, MI 48109, USA}
\affiliation[z]{University of Michigan, 500 S. State Street, Ann Arbor, MI 48109, USA}
\affiliation[aa]{Department of Physics, The University of Texas at Dallas, 800 W. Campbell Rd., Richardson, TX 75080, USA}
\affiliation[ab]{NSF NOIRLab, 950 N. Cherry Ave., Tucson, AZ 85719, USA}
\affiliation[ac]{Department of Physics and Astronomy, University of California, Irvine, 92697, USA}
\affiliation[ad]{The Ohio State University, Columbus, 43210 OH, USA}
\affiliation[ae]{Departament de F\'{i}sica, Serra H\'{u}nter, Universitat Aut\`{o}noma de Barcelona, 08193 Bellaterra (Barcelona), Spain}
\affiliation[af]{Department of Physics and Astronomy, University of Waterloo, 200 University Ave W, Waterloo, ON N2L 3G1, Canada}
\affiliation[ag]{Perimeter Institute for Theoretical Physics, 31 Caroline St. North, Waterloo, ON N2L 2Y5, Canada}
\affiliation[ah]{Waterloo Centre for Astrophysics, University of Waterloo, 200 University Ave W, Waterloo, ON N2L 3G1, Canada}
\affiliation[ai]{Instituto de Astrof\'{i}sica de Andaluc\'{i}a (CSIC), Glorieta de la Astronom\'{i}a, s/n, E-18008 Granada, Spain}
\affiliation[aj]{Departament de F\'isica, EEBE, Universitat Polit\`ecnica de Catalunya, c/Eduard Maristany 10, 08930 Barcelona, Spain}
\affiliation[ak]{Department of Physics and Astronomy, Sejong University, 209 Neungdong-ro, Gwangjin-gu, Seoul 05006, Republic of Korea}
\affiliation[al]{CIEMAT, Avenida Complutense 40, E-28040 Madrid, Spain}
\affiliation[am]{Department of Astronomy, Tsinghua University, 30 Shuangqing Road, Haidian District, Beijing, China, 100190}
\affiliation[an]{National Astronomical Observatories, Chinese Academy of Sciences, A20 Datun Road, Chaoyang District, Beijing, 100101, P.~R.~China}

\abstract{We constrain local primordial non-Gaussianity by a combined analysis of auto and cross-correlations of DESI DR1 tracers, leveraging LRGs and QSOs as well as ELGs between $0.8<z<3.1$. By cross-validating the signal across different clustering tracers within the same redshift range, we evaluate potential systematics in the $\fnl$ measurements, capitalizing on the reduced susceptibility of cross-correlations to non-common systematics.  
We find that the cross-correlation between LRG and quasars can robustly improve the DESI DR1 $\fnl$ constraints, by $\sim9\%$ to a measurement of $\fnl=2.1_{-8.3}^{+8.8}$ at 68\% confidence. 
\hl{On the other hand, we do not find a clear improvement when including the DESI DR1 ELG sample. Mock tests predict an additional $\sim8\%$ gain with statistical scatter, and the lack of improvement in the data remains consistent with this expectation.}
This project serves as an exploratory analysis of DESI ELG clustering for $\fnl$ through its cross-correlation in preparation for future DESI data analyses.
}

\begin{document}
% \linenumbers
\maketitle
\flushbottom

% BODY OF PAPER 

\section{Introduction}
\label{sec:introduction}
Inflation is the leading theoretical model describing the rapid expansion at early times of the universe. It addresses the issues left unresolved by the Hot Big Bang, namely, the flatness problem, the horizon problem, and the monopole problem \cite{Guth_1981, Linde_1982, Albrecht_1982}. Although inflation successfully explains these features, the exact nature of the fields responsible for this exponential expansion remains unknown.
During this early epoch, the universe was opaque, and thus direct observations from this time are not available to us. However, we can probe the inflationary dynamics through imprints in the cosmic microwave background \cite{Seljak_1997,Zaldarriaga_1997} and large-scale structure, such as primordial non-Gaussianity (PNG) \cite{planckcollaboration2019planck, Cagliari_2024}. The amplitude of PNG of the local type is parameterized by the non-linear coupling constant $\fnl$,
\begin{equation}
   \Phi(\mathbf{x}) = \phi(\mathbf{x}) + \fnl \left(\phi^2(\mathbf{x}) - \langle \phi^2 \rangle \right) 
\end{equation}
where $\Phi$ is the primordial potential and $\phi$ is assumed to be a Gaussian random field \cite{Matarrese_2000,Komatsu_2001}.
In particular, a non-zero detection with precision of $\mathcal{O}(1)$ of local-type PNG \hl{would rule out single-field infation} \cite{Paolo_Creminelli_2004,chen_2010_review}.

The best constraint on  $\fnl$ at the moment is from the Planck collaboration \cite{PR3}, using measurements from the CMB bispectrum, $\fnl=-0.9 \pm 5.1$ at 68\% confidence \cite{planckcollaboration2019planck}. 
However, this measurement is cosmic variance limited, since it is confined to the two-dimensional data at the last scattering surface. \hl{Hence, we do not expect significant improvements in precision from CMB data \cite{planckcollaboration2019planck}.}

On the other hand, three-dimensional galaxy clustering allows us to push for better precision by expanding the survey cosmic volume to higher redshift and tapping into the statistical power within 3D galaxy clustering.
This entails constraining the $\fnl$ signature by measuring the scale-dependent bias effect on the large-scale clustering of biased tracers \cite{Slosar_2008,dalal2008imprints}.
Before DESI, the best constraint from LSS was obtained with the QSO sample from the 16th data release of the extended Baryon Acoustic Oscillation Spectroscopic Survey (eBOSS) \cite{Ross.eBOSS.2020,eBOSS.DR16} which yielded $-23<\fnl<21$ at 68\% confidence \cite{Cagliari_2024}.
The latest best constraint from the LSS is from the joint constraint of DESI DR1 LRGs and QSOs \cite{ChaussidonY1fnl}, which reported $\fnl=-3.6^{+9.0}_{-9.1}$ at 68\% confidence.

The DESI DR1 ELGs are not included in the latest DESI DR1 constraint of $\fnl$. This is because their imaging systematics dominate at large scales \cite{Rosado-Marín_2025,KP3s15-Ross}, the scales most relevant for $\fnl$ analysis, by many orders of magnitude compared to the level of the $\fnl$ signal expected based on the CMB constraint. The complex and nonlinear nature of their systematics, coupled with their large magnitude, raises the risk of either residual effects, even after rigorous mitigation efforts, or overcorrection of the true clustering signal due to the large flexibility we allow during the mitigation. The low galaxy bias of the ELGs makes its $\fnl$ signal intrinsically weak, making them particularly prone to these systematic effects. 

In this paper, we propose and validate ways to include the information of ELGs in the $\fnl$ analysis in a way that is less prone to the residual systematics. We include all cross-correlations with ELGs as well as all other possible cross-correlations of DESI DR1 tracers, in addition to the auto clustering of LRGs and QSOs. By utilizing the cross-correlations, we are limiting the effect of potential residual systematics in ELGs, as \hl{non-common} systematics affecting different tracers will \hl{be uncorrelated}. By simultaneously constraining $\fnl$ using auto as well as all possible cross-correlations of the DESI DR1 tracers, we also seek to improve the precision. If there are any residual systematics, unless the effect is universal for all tracers, one can imagine the $\fnl$ constraint will differ depending on the different combinations of auto and cross-correlations across different tracers. We test such dependence to inspect and quote a systematic error on our $\fnl$ constraint.

This work is organized as follows. In \cref{sec:data} we describe the data samples used in our analysis. In \cref{sec:methods} we go over the measurement of the power spectrum; the theory model used to measure the scale dependent bias from large-scale structure; known integral constraints that impact the power spectrum; and detail the configuration of the parameter estimation. In \cref{sec:results} we discuss our results, including our expectation from the mocks and findings from the data. Furthermore, we discuss our findings from the data in the context of the mocks. Finally, in \cref{sec:conclusion} we summarize our findings and discuss what is left for future investigation.

\section{Data}
\label{sec:data}
The DESI instrument \cite{DESI2022.KP1.Instr} is a multi-fiber spectroscopic instrument mounted on the Nicholas U. Mayall Telescope at Kitt Peak, Arizona. Its 5,000 optical fibers are controlled by a corresponding number of robotic positioners to simultaneously obtain spectra for 5,000 astronomical `targets' \cite{TS.Pipeline.Myers.2023} within the 7 square degree field of view of the focal plane \cite{Corrector.Miller.2023,FocalPlane.Silber.2023,DESIfa}. The fibers are arranged into ten `petals', and every group of targets assigned to a set of 5,000 fibers, namely a `tile', is associated with a central sky coordinate.

DESI’s main survey began on May 14, 2021, following an initial survey-validation phase \cite{DESI2023a.KP1.SV}. Our analysis uses the main-survey observations included in DESI DR1 \cite{DESI2024.I.DR1}, covering data taken through June 14, 2022. The version of the spectroscopic redshifts for DESI DR1, denoted as `iron', were reduced using the DESI spectroscopic pipeline \cite{Spectro.Pipeline.Guy.2023}.\footnote{Processed with version 23.1 of the DESI software, available on NERSC via source /global/common/software/desi/desi\_environment.sh 23.1.} 
In \cite{KP3s15-Ross} and \cite{DESI2024.II.KP3} they describe the generation of the LSS catalogs and how they are processed for the DESI DR1 \cite{desiDR1}. 
\hl{With this wealth of data the DESI collaboration presented measurements of baryon acoustic oscillation (BAO) \cite{DESI2024.III.KP4, DESI2024.IV.KP6, DESI2024.VI.KP7A,DESIDR2.BAO} and redshift space distortions (RSD) \cite{DESI2024.V.KP5,DESI2024.VII.KP7B}.}
Particularly important details on how we use the data are described in the following subsections.

\subsection{DR1 Data Samples}
\label{subsec:dr1_data}
We extend our analysis from what was presented in \cite{ChaussidonY1fnl} by including ELGs and their cross-correlations, also the cross-correlation between LRGs and QSOs. In \cite{Rosado-Marín_2025} they showed that even after mitigation of imaging systematics, the DESI DR1 ELG sample shows significant excess clustering at the scales relevant for constraining $\fnl$. Therefore, \hl{we include ELGs in an $\fnl$ analysis} through their cross-correlations with QSOs and LRGs. These cross-tracer samples serve as different tracers of the same underlying $\fnl$ signature, and are especially useful for checking systematics, as we do not expect any bias from non-common additive systematics in the cross-correlations. 

Throughout this work we use version v1.5 of the DESI DR1 LSS catalogs\footnote{\url{https://data.desi.lbl.gov/doc/releases/dr1/}} \cite{desiDR1}. Let us first define the auto-tracer samples:
\begin{itemize}
    \item Luminous red galaxies (LRGs): The LRG sample consists of 2,138,627 good redshift objects within $0.4 < z < 1.1$, covering $5,740\,\mathrm{deg}^2$. Imaging systematics are mitigated with the linear regression described in \cite{DESI2024.II.KP3}.

    \item Emission line galaxies (ELGs): The ELG sample consists of 2,432,072 good redshift objects within $0.8 < z < 1.6$, covering $5,924\,\mathrm{deg}^2$. Imaging systematics are mitigated using \sysnet as described in \cite{Rosado-Marín_2025}
    
    \item Quasars (QSO): The QSO sample consists of 1,190,839 good redshift objects within $0.8 < z < 3.1$, covering $7,056\,\mathrm{deg}^2$. The extended redshift range for this QSO sample follows the DESI DR1 PNG analysis in \cite{ChaussidonY1fnl}, and is expected to be further extended to $z=3.5$ for the DESI DR2 PNG analysis. Imaging systematics are treated using linear weights derived using \regressis as defined in \cite{ChaussidonY1fnl}.
\end{itemize}
Now we define the cross-tracer samples, which are measured at the redshift range at which the tracers overlap.
\begin{itemize}
    \item LRGxQSO (\lxq): This sample consists of 1,003,668 objects (859,822 LRGs; 143,846 QSOs) and represents the cross-correlation between the LRGs and the QSOs within $\zrlrgxqso$. 
    \item ELGxLRG (\exl): This sample consists of 1,876,187  objects (1,016,365 ELGs; 859,822 LRGs) and represents the cross-correlation between the ELGs and the LRGs within $\zrelgxlrg$.
    \item ELGxQSO (\exq): This sample consists of 2,934,693  objects (2,4320,72 ELGs; 502,621 QSOs) and represents the cross-correlation between the ELGs and the QSOs within $\zrelgxqso$.
\end{itemize}

\subsection{Weighting DESI DR1 Catalogs for $\fnl$}

DESI DR1 catalogs contain weight columns that assign corrections for various known completeness and systematics effects at a per-object level\footnote{\hl{The resolution of the imaging systematics weights is not at the object level, they are assigned from pixel maps with HEALPix resolution of  $N_\mathrm{side} = 256$.}}. The totality of the weights used in this work, in the absence of optimal weighting schemes, are contained within
\begin{equation}
\label{eq:weights_eq}
    w_\mathrm{tot} = w_\mathrm{sys}\times w_\mathrm{comp}\times w_\mathrm{zfail},
\end{equation}
where $w_\mathrm{sys}$ corrects for the spurious correlations between the target density field and known imaging systematics \cite{DESI2024.II.KP3,Rosado-Marín_2025,ChaussidonY1fnl}, $w_\mathrm{comp}$ accounts for fiber-assignment incompleteness variations in the sample \cite{DESI2024.II.KP3}, and $w_\mathrm{zfail}$ accounts for changes in the relative redshift success rate \cite{DESI2024.II.KP3}.

The imaging systematics weights of the samples considered for this analysis are contained within $w_\mathrm{sys}$ and are generated differently for each tracer. Starting with the LRGs, the imaging systematics are mitigated with the linear regression scheme described in \cite{DESI2024.II.KP3}\hl{, which fits the data to the binned statistics of imaging systematic templates.} The ELGs are mitigated using the \sysnet pipeline as described in \cite{Rosado-Marín_2025}\hl{, which uses a fully connected Feed Forward Neural Network to model the relationship between imaging systematics and target density}. The QSOs used for the auto-correlation are treated using linear weights derived using \regressis as defined in \cite{ChaussidonY1fnl}\hl{, which fits the data to the fluctuations of imaging systematic templates at the \textsc{HEALPix} level}.
\hl{The QSOs} used in the cross-correlated samples are using a set of linear weights that are derived using the same linear regression pipeline employed for the LRGs but regressing against the imaging systematics relevant for QSOs and their corresponding photometric regions (i.e. North, DES, and SnoDES\hl{; as done in \cite{ChaussidonY1fnl}}). These linear weights for the QSOs are not directly available in the v1.5 clustering catalogs, but are made accessible through the `full' catalogs. Only to assign them to the data and randoms clustering catalogs via the \texttt{TARGETID} column.\footnote{ \hl{The derived imaging systematics weights for auto- and cross-QSOs are computed with different pipelines but the model used in the regression is linear for both cases, so we assume that the difference between regression pipelines is negligible.}}

In addition to the weights described in \cref{eq:weights_eq}, we also implement optimal weighting schemes, namely FKP weights and optimal quadratic estimator (OQE) weights \hl{for the scale-dependent bias. The FKP weights optimally weight a galaxy sample such that the expected variance of the measured power spectrum at the scale of interest is minimized \cite{Feldman_1994}. 
The OQE weights further enhance the FKP weights by accounting for the redshift evolution of the scale-dependent bias signature in the measured power spectrum \cite{Castorina_2019}.}  
In \cite{DESI2024.II.KP3} the FKP weights, $\wfkp$, were computed with the fixed power spectrum amplitude, $P_0$, evaluated at the optimal $k$ for BAO analysis. However, for constraining PNG, the relevant scales are around $k=0.01\,\hmpc$. Hence, we use $P_0=5\times10^4\,\mpchcubed$ for LRGs, $P_0=2\times10^4\,\mpchcubed$ for ELGs, and $P_0=3\times10^4\,\mpchcubed$ for QSOs. \hl{The auto-QSO sample is the only OQE weighted sample in our analysis, in \cite{ChaussidonY1fnl} they found that the OQE weights are most important for this sample.}

\begin{table}
    \centering
    \caption{Summary of the sample definitions and their relevant quantities. From left to right, we define: the redshift ranges, the effective redshifts, the fixed power spectrum amplitudes (used when computing FKP weights), and the minimum and maximum wavenumber used when constraining $\fnl$ for each sample. The redshifts used in the $\zeff$ computation are weighted to correct for completeness, redshift failures, and imaging systematics, as described in \cref{eq:weights_eq}. Except for the auto-QSO, we use only the monopole of the power spectrum and use FKP weights. Therefore, the per multipole $\zeff$ of the QSO auto-correlation sample is weighted by the OQE weights instead of the FKP weights, as indicated by the OQE superscript.}
    \begin{tabular}{ccccccc}\toprule
    & & & $z^{\mathrm{OQE}}_{\mathrm{eff}}$  & &\multicolumn{2}{c}{$k\,[\hmpc]$}\\
         &  Redshift &  $z^{\mathrm{FKP}}_\mathrm{eff}$ &  ($\ell=0$/$\ell=2$) & $P_0\,[\mpchcubed]$& $k_\mathrm{min}$ & $k_\mathrm{max}$  \\\midrule
         LRG&  $0.4<1.1$& 0.737 & -& $5\times10^4$& $6\times10^{-3}$& $8\times10^{-2}$\\
         ELG& $0.8<1.6$& 1.186 &- &$2\times10^4$ & $1\times10^{-2}$& $8\times10^{-2}$\\ 
         QSO&  $0.8<3.1$&  1.662 & 2.094 / 2.001& $3\times10^4$& $3\times10^{-3}$& $8\times10^{-2}$ \\
         LRGxQSO&  $0.8<1.1$&  0.956 & -  &- & $6\times10^{-3}$& $8\times10^{-2}$ \\
         ELGxLRG&  $0.8<1.1$&  0.918 & -  &- & $6\times10^{-3}$& $8\times10^{-2}$ \\
 ELGxQSO& $0.8<1.6$& 1.203 &- &- & $6\times10^{-3}$& $8\times10^{-2}$ \\\bottomrule
    \end{tabular}  
    \label{tab:zeff}
\end{table}

In \cref{tab:zeff} we summarize what has been discussed until now, i.e., the redshift range, effective redshift, maximum amplitude of the power spectrum evaluated at $k=0.01\,\hmpc$ used for computing the FKP weights, and the minimum and maximum wavenumber used when constraining $\fnl$ for each sample (discussed in \cref{subsec:parameter_estimation}). The values shown in this table are used consistently throughout this paper unless stated otherwise.

\subsection{EZmocks}
\label{subsec:EZmocks}
We use EZmocks (effective Zel'dovich aproximation mocks) \cite{Chuang_2014} to construct the covariance matrices and perform validation tests that recover unbiased $\fnl$ measurements and estimate the expected gain from different sample combinations. EZmocks are a suite of fast simulations that use the Zel'dovich approximation \cite{Zeldovich_1970} to generate the dark matter density field at a given redshift and are calibrated with clustering statistics from observations or N-body simulations. The description of the generation and calibration of DESI DR1 EZmocks is given in \cite{KP3s8-Zhao}. 
Our analysis does not require using N-body simulations, as we are interested in the large scale expected signal and noise, where the Zel'dovich approximation is valid.
\hl{However, it is important that the simulation boxes are large enough to cover the full DESI survey volume without duplications. The EZmocks were generated from $6\,h^{-1}\mathrm{Gpc}$ boxes, with 1000 boxes per galactic cap (NGC and SGC), more than enough for our purpose.}

\hl{The SecondGen EZmocks generated for DESI DR1 LRGs, QSOs, and ELGs have the estimated fiber assignment effect from the Fast Fiber Assignment (FFA) \cite{Bianchi_2025} method. The QSOs of these FFA mocks only extend to $z=2.1$, and the PNG constraint is stronger the more large-scale modes we probe. Therefore, we also use the QSOs from the FirstGen EZmocks mocks generated in \cite{ChaussidonY1fnl} that reach $z=3.1$. However, the mocks in \cite{ChaussidonY1fnl} do not include the fiber assignment effect. We assume the impact of ignoring the fiber assignment for QSO is negligible since QSOs have the priority in fiber assignment \cite{Bianchi_2018, ChaussidonY1fnl, Bianchi_2025}.}

\hl{We now make an important distinction between the QSOs from FFA EZmocks ($0.8<z<2.1$) and the QSOs from the mocks in \cite{ChaussidonY1fnl} ($\zrqso$). The former are used for measuring the cross-power spectrum of the LRGxQSO and ELGxQSO samples since they are computed at their overlapping redshift range; while the latter are used for measuring the power spectrum of the $\zrqso$ QSOs.}

The DESI DR1 EZmocks for LRGs and ELGs are stitched from two snapshots, while the QSOs are generated using a single snapshot. The LRGs were constructed from a $z=0.5$ snapshot for $0.4 < z < 0.6$ and a $z=0.8$ snapshot for $0.6 < z < 1.1$. The ELGs were generated from a $z=0.95$ snapshot for $0.8 < z < 1.1$ and a $z=1.325$ snapshot for $1.1 < z < 1.6$. Finally, the QSOs are generated from a $z=1.4$ snapshot for $0.8 < z < 2.1$ and $0.8 < z < 3.1$. 

We use 1000 mocks per tracer to construct the covariance matrices for the auto-tracer samples and cross-tracer samples and the cross-covariances among them, except for the auto-QSO covariance.
\hl{The covariance matrix of the auto-QSO is generated from the mocks with no FFA from \cite{ChaussidonY1fnl} hence we ignore all cross-covariance between the auto-QSO and other tracers, which is a reasonable approximation due to the high shot noise of the QSO sample.}
The covariances of cross-tracers with QSO, e.g., LRGxQSO, are generated using the FFA EZmocks, which extend to redshift $z=2.1$, enough for computing the cross-power spectra within the overlapping redshifts.

\begin{figure}
    \centering
    \includegraphics[width=\linewidth]{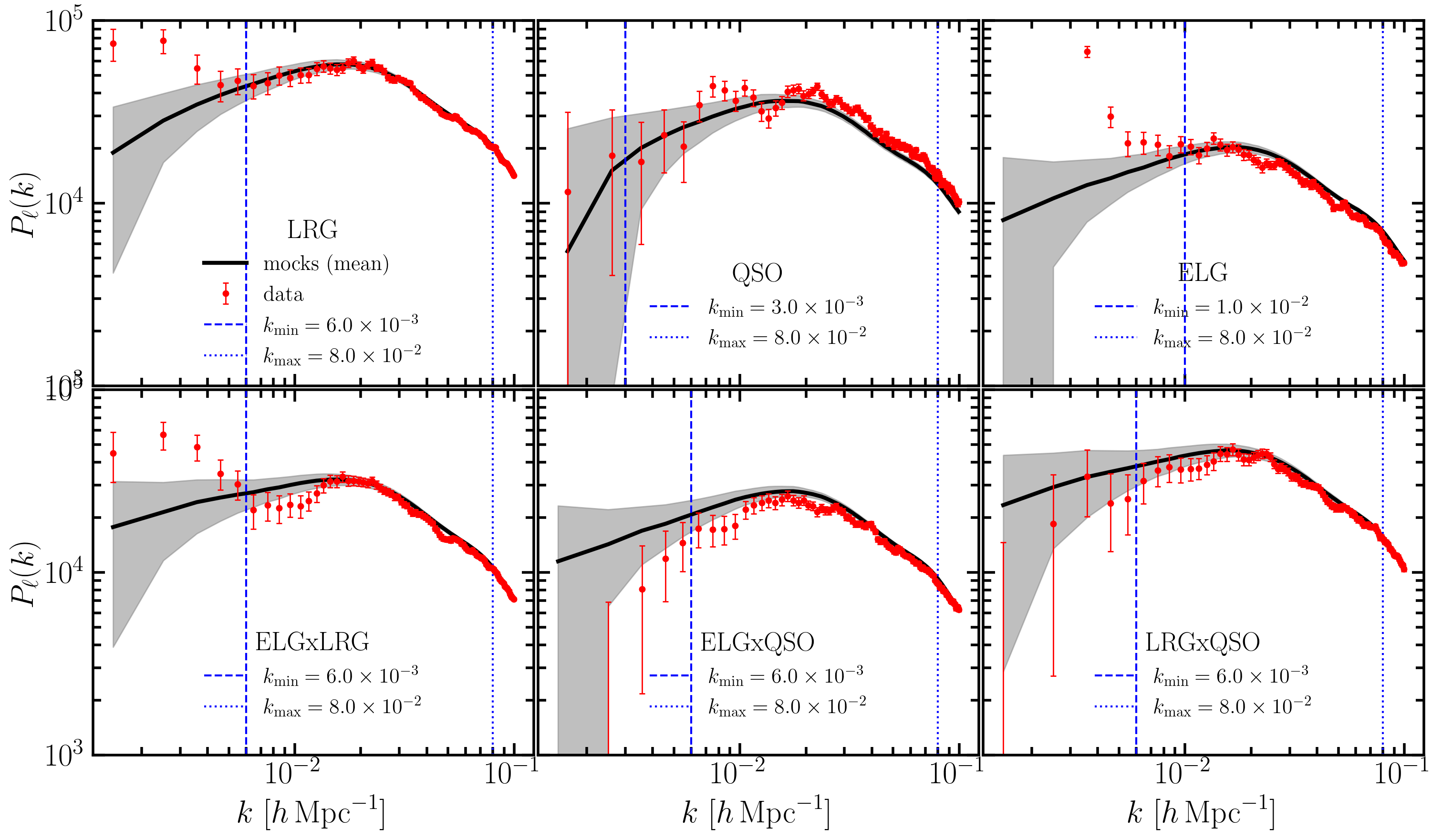}
    \caption{Two-point galaxy clustering of DR1 DESI tracers in Fourier space. The upper panels show the auto-power spectrum for LRGs, QSOs, and ELGs from left to right.
    The lower panels show the cross-power spectrum for the cross-correlation between DESI tracers; \elgxlrg, \elgxqso, \lrgxqso\, from left to right. The black curves represent the mean of 1000 EZmocks power spectrum, while the grey shaded region is the $\pm1\sigma$ region from these mocks. The red markers show the power spectrum measured from the DR1 DESI data with imaging weights applied. We also highlight the minimum and maximum wavenumber used when running the $\fnl$ pipeline with vertical dashed and dotted blue lines, respectively.}
    \label{fig:power_panels}
\end{figure}

In \cref{fig:power_panels} we show the two-point galaxy clustering in Fourier space measured from the mean of 1000 EZmocks (solid black curve) and the DR1 data (red markers). The grey shaded area is the $\pm 1\sigma$ region from the mocks. The power spectrum from the EZmocks matches the observed DR1 clustering in shape but not in amplitude. This mismatch in amplitude was observed in \cite{ChaussidonY1fnl} and is due to the EZmocks being generated at different redshift snapshots than the effective redshift of the data. 
Hence, the bias is different from that expected from the observed data.

We handle the mismatch of the power spectrum by renormalizing the bias of the EZmocks to match the amplitude of the observed power spectrum, as in \cite{ChaussidonY1fnl}. For the auto-tracer samples, we first measure the bias, $b$, of the mocks and compute the bias expected from the data, $b_{\mathrm{exp}}$. Then we use the ratio $b_{\mathrm{exp}}/b$ and the RSD corrections to scale the amplitude of the EZmock power spectrum mulitpoles by
\begin{equation}
    \begin{cases}
    \begin{array}{ll}
    P_0(k)\xrightarrow[]{} \frac{b^2_{\mathrm{exp}}}{b^2} \frac{1+2/3\beta_{\mathrm{exp}}+1/5\beta^2_{\mathrm{exp}}}{1+2/3\beta+1/5\beta^2} \times P_0(k)\\
    P_2(k)\xrightarrow[]{} \frac{b^2_{\mathrm{exp}}}{b^2} \frac{4/3\beta_{\mathrm{exp}}+4/7\beta^2_{\mathrm{exp}}}{4/3\beta+4/7\beta^2} \times P_2(k)
    \end{array}
    \end{cases}
    ,
\end{equation}
where $\beta=f(z_\mathrm{eff})/b$. On the other hand, for the cross-tracer samples we compute the renormalization factors for each tracer at the effective redshift of the cross-tracer sample and use their product to rescale the cross-power spectrum. 

\section{Analysis methods}
\label{sec:methods}
In this section we describe how we measure and model the power spectrum in our analysis, and account for the radial and angular integral constraints that impact our measured power spectrum. We also describe the estimated parameters and the generation of the data vectors used throughout the analysis.

\subsection{Measuring the Power Spectrum}
In this work, we use the power spectrum measured from the LSS catalogs as our observables. The power spectrum estimator and code (\textsc{pypower}\footnote{\url{https://github.com/cosmodesi/pypower}}) are described in \cite{DESI2024.II.KP3}. All power spectra use 10 random catalogs \hl{with a density of $2,500/\mathrm{deg}^2$ each}, and are computed with a physical box size of $6000\,h^{-1}\rm Mpc$ with a grid cell size of $6\,h^{-1}\rm Mpc$.
When computing the cross-power spectrum, the FFT of the weighted density field of the first tracer is multiplied by the complex conjugate of the FFT of the field of the second tracer.

\subsection{Power Spectrum Model}

In Fourier space, the initial gravitational field is related to the matter overdensity field, $\delta_m$, by $\delta_m(k,z) = T_{\Phi\rightarrow\delta}(k,z)\Phi(\mathbf{k})$,
where $T_{\Phi\rightarrow\delta}(k,z)$ is the transfer function between the primordial gravitational potential, $\Phi$, and the matter overdensity. We can calculate $T_{\Phi\rightarrow\delta}(k,z)$ with \textsc{cosmoprimo}\footnote{\url{https://github.com/cosmodesi/cosmoprimo}} by:
\begin{equation}
    T_{\Phi\rightarrow\delta}(k,z) = \sqrt{\frac{P_\mathrm{lin}(k,z)}{P_\Phi(k)}},
\end{equation}
where $P_\mathrm{lin}(k,z)$ is the linear matter power spectrum and $P_\Phi(k)$ is the primordial potential power spectrum.

Primordial non-Gaussianity as parametrized by $\fnl$ introduces a scale dependent bias \cite{Slosar_2008,dalal2008imprints}, $\Delta b(k,z)$, given by 
\begin{equation}
\label{eq:scale_dep_bias}
     \Delta b(k,z) \equiv  \frac{b_\phi}{T_{\Phi\rightarrow\delta}(k,z)}f_\mathrm{NL} \propto \frac{b_\phi}{k^2T_{\Phi\rightarrow\Phi}(k,z)}\fnl. 
\end{equation}
The relation $T_{\Phi\rightarrow\delta}(k,z)\propto k^2T_{\Phi\rightarrow\Phi}(k,z)$ contains the scale-dependent bias, with $k^2$ arising from solving Poisson equation. In addition, we use $b_\phi(z) = 2\delta_c(b_1(z)-p)$ where $b_1$ is the bias of the tracer and $p$ quantifies the merger history of the tracer. The $\delta_c = 1.686$ is the critical density in the spherical collapse model in a Einstein-deSitter Universe (zero curvature matter dominated universe).
\hl{Notice that $\fnl$ is degenerate with the bias coefficient $b_\phi$, so we typically assume a value for $p$. It is known that this this relationship does not hold in hydrodynamic simulations, and the determination of $b_\phi$ is being actively studied \cite{Biagetti_2017,Barreira_2020a,Barreira_2020b,Sullivan_2023,Boryana_2024,Gutierrez_2024,perez2026impactgalaxyformationgalaxy}.
Thus, in this work we assume a recent merger history for QSOs, $p=1.6$ \cite{Slosar_2008}, and the mass universality relation, $p=1$ \cite{dalal2008imprints}, for LRGs. For ELGs, we fix $p=1.6$. Internal tests were performed, and fixing $p=1$ for ELGs yields consistent results.
We note that \cite{fondi2026assemblybiaslocalprimordial} recently found a physically motivated value of $p=1.4$ for DESI QSOs, and in \cite{Barreira_2020a,Barreira_2020b} they find that $p=0.55$ is preferred by stellar-mass selected samples. We present our results with these updated values of $p$ in \cref{subsec:results_comparisons}.}

The relation between the tracer overdensity, $\delta_g$, and matter overdensity with scale dependent bias by PNG is 
\begin{equation}
    \delta_\mathrm{g} = \left[b_1 +  \Delta b + f\mu^2 \right] \delta_m,
\end{equation}
where $f=d\ln D/d\ln a$ is the logarithmic growth rate, and $\mu$ is the cosine of the angle between the Fourier modes and the line of sight.

If we use the power spectrum as the observable to probe PNG, then we need to construct a model for the power spectrum that factors in the tracer bias scale dependence due to PNG. Using linear theory to predict the power spectrum of the biased tracer in redshift space is 
\begin{equation}\label{multi-model}
    P_{ij}(k,\mu) = G_i(k,\mu;\Sigma_{s,i})G_j(k,\mu;\Sigma_{s,j}) \left[b_{1,i} +  \Delta b_{1,i} +f\mu^2 \right] \left[b_{1,j} +  \Delta b_{1,j} +f\mu^2 \right] P_m(k) + \textrm{sn}_{0,ij},
\end{equation}
where the suffixes $i$ and $j$ are used to represent different biased tracers. Also, $\textrm{sn}_{0}$ is a free parameter for any residual shotnoise, $G(k,\mu;\Sigma_s) = \left[ 1 + (k\mu\Sigma_s)^2/2 \right]^{-1}$
accounts for the Finger-of-God (FoG) effect in redshift space, and $\Sigma_s$ is the dampening velocity dispersion.

\subsection{Integral Constraints}

The so-called radial and angular integral constraints \cite{de_Mattia_2019} impact the measured power spectrum, especially at large scales. Fortunately, their impact can be estimated from mocks and their contributions folded into the window matrix that convolves our model \cite{ChaussidonY1fnl}.

To ensure that the random catalogs have the same redshift distribution as the data, a method called \emph{shuffling} is employed \cite{Ross_2012}. This method subsamples the redshifts of the data catalogs and assigns them to the random catalogs. However, assigning redshifts to randoms in this way removes radial modes in the measured power spectrum. This radial mode removal effect is known as the radial integral constraint (RIC) \cite{de_Mattia_2019}. We quantify and account for this effect as in \cite{ChaussidonY1fnl}. They used the mean of the power spectra of 50 EZmocks with and without shuffling applied and computed the contribution of the RIC on the window function. In this way, they could apply a multiplicative correction of the RIC. \hl{In this paper, the constraints performed on the DR1 data and the EZmocks with FFA contain the appropriate RIC contribution.}

The other integral constraint we take into account is the angular integral constraint (AIC). A flavor of this effect is introduced when mitigating imaging systematics. 
This results in the removal of large scale modes in the power spectrum, biasing our $\fnl$ measurement. We estimate the AIC contribution in a similar manner as done for the RIC, following \cite{ChaussidonY1fnl}. We apply the same mitigation schemes used for the data on the first 50 EZmocks, which are null for imaging systematics, and compute the contribution of the AIC on the window function. 

\subsection{Parameter estimation}
\label{subsec:parameter_estimation} 

\begin{table}
    \centering
    \caption{Priors used for the free parameters in the model. Here $\fnl$ is the measured amplitude of non-Gaussianity, which is the universal fitting parameter, $b_{1,i}$ is the linear galaxy bias of the of tracer $i$, $\Sigma_{s,i}$ is the dampening velocity dispersion, and $\textrm{sn}_{0,i}$ is the shotnoise which we marginalize over. When adding cross tracer clustering, we introduce an additional shot noise parameter $\textrm{sn}_{0,ij}$ for the \hl{cross-correlation of tracers $i$ and $j$}. Therefore, the total number of fitting parameters $n_p=1+N_{\rm auto}\times 3+N_{\rm cross}$, where $N_{\rm auto}$ and $N_{\rm cross}$ are the number of auto- and cross-tracer samples, respectively.}
    \begin{tabular}{cl}\toprule
         Parameter& Prior\\\midrule
         $f^{\mathrm{loc}}_\mathrm{NL}$& $\mathcal{U}(-300,300)$\\\midrule
         $b_{1,i}$& $\mathcal{U}(0.1,10)$\\
         $\Sigma_{s,i}$& $\mathcal{U}(0,10)$\\
         $\textrm{sn}_{0,i}$& $\mathcal{N}(0,1000)$\\ \midrule
         $\textrm{sn}_{0,ij}$& $\mathcal{N}(0,1000)$\\\bottomrule
    \end{tabular} 
    \label{tab:priors}
\end{table}

In \cref{subsec:dr1_data} we established the data samples used in our analysis. Here we define the data vectors for constraining $\fnl$ and their corresponding limits on $k$ (which are also shown in \cref{tab:zeff} and as vertical lines in \cref{fig:power_panels}): 
\begin{itemize}
    \item LRG: This sample is FKP weighted and is composed of LRGs within $0.4<z<1.1$, as determined in \cite{ChaussidonY1fnl} we limit the monopole to $k_\mathrm{min}=6\times10^{-3}\hmpc$ and $k_\mathrm{max}=8\times10^{-2}\hmpc$ with $\Delta k=1\times10^{-3}\hmpc$. 
    
    \item QSO: This sample is OQE weighted and is composed of QSOs within $0.8<z<3.1$, as determined in \cite{ChaussidonY1fnl} we limit the monopole and quadrupole to $k_\mathrm{min}=3\times10^{-3}\hmpc$ and $k_\mathrm{max}=8\times10^{-2}\hmpc$. For QSOs we use $\Delta k=1\times10^{-3}\hmpc$ and $\Delta k=2\times10^{-3}\hmpc$ for the monopole and quadrupole, respectively. 
    
    \item ELG:  This sample is FKP weighted and is composed of ELGs within $0.8<z<1.6$, we limit the monopole to $k_\mathrm{min}=1\times10^{-2}\hmpc$ and $k_\mathrm{max}=8\times10^{-2}\hmpc$ with $\Delta k=1\times10^{-3}\hmpc$. We make a conservative choice, at the cost of precision, and cut the ELGs to $k_\mathrm{min}=1\times10^{-2}\hmpc$. In this way, we do not worry about the residual systematics in the observed ELG power spectrum biasing our $\fnl$ measurement. \cref{fig:power_panels} shows $k_\mathrm{min}$ with a vertical dashed blue line, notice that the data removed by this choice of $k_\mathrm{min}$ would have positively biased the constraints on $\fnl$.

    \item LRG+QSO: This case combines the LRG and QSO auto-clustering data as described above. The covariance matrix has no cross-covariance information between the LRG and the QSO sample, since the EZmocks for QSOs were made independently of the rest of the tracers. The cross-covariance between LRG and QSO will be included in future studies. \hl{We note that the effective volumes of the QSO samples in the ranges $0.8 < z < 3.1$ and $0.8 < z < 1.1$ are $8.4\,(\mathrm{Gpc}/h)^3$ and $1.1\,(\mathrm{Gpc}/h)^3$, respectively. Consequently, the large-scale pairs (or low $k$ modes) in the auto-QSO sample ($0.8 < z < 3.1$) are largely distinct from those contributing to the QSO and LRG pairs (or $k$ modes) in the LRGxQSO sample ($0.8 < z < 1.1$). Therefore, the covariance between the auto-QSO and LRGxQSO measurements, which is the added sample most relevant to our analysis, is expected to be small.}
    
    \item L+Q+LxQ: This case adds the LRGxQSO cross-power and follows the same limits in $k$ discussed for the LRG sample. The covariance matrix contains cross-covariance information between the LRG and the LRGxQSO samples. 
    
    \item L+Q+LxQ+E: This case adds the ELG auto-clustering data as described above. The covariance matrix contains cross-covariance information between the LRG, LRGxQSO, and ELG samples. \hl{The inclusion of the ELG auto-clustering data is not to constrain $\fnl$ but to constrain the ELG bias when including the ELGxLRG and ELGxQSO in the case below.}
    
    \item L+Q+LxQ+ExL+ExQ+E: This case adds the ELGxLRG and ELGxQSO cross-power and follows the same limits in $k$ discussed for the LRG sample. The covariance matrix contains cross-covariance information between the LRG, LRGxQSO, ELGxLRG, ELGxQSO, and ELG samples.
\end{itemize}

\hl{As summarized in \cref{tab:priors}, the priors used for each parameter are: $\mathcal{U}(-300,300)$ for $\fnl$, $\mathcal{U}(0.1,10)$ for $b_1$, $\mathcal{N}(0,1000)$ for $s_{n,0}$, and $\mathcal{U}(0,10)$ for $\Sigma_s$. The only universal parameter is $\fnl$, while $b_1$ and $\Sigma_s$ are shared across tracers of the same type. Residual shot noise, on the other hand, is considered to be unique for each sample (samples were defined in \cref{subsec:dr1_data}). In addition, we fixed the assembly bias parameter $p$ for each tracer to $p=1.6$ for QSOs and ELGs, and $p=1$ for LRGs.}

\hl{We account for the linear bias redshift evolution described in \cite{Laurent_2017,ChaussidonY1fnl} by}
\begin{equation}
    b_{i}(z)=\alpha(1+z)^2+\beta_i,
\end{equation}
\hl{where the subscript $i$ indicates that the value corresponds to tracer $i$. 
However, the effective redshifts are different for the auto-correlation and cross-correlation. Hence, we link the linear biases of the cross-tracers within the cross-correlated samples to their corresponding auto-correlated samples. 
For example, if we represent the linear bias of tracer $i$ within the cross-correlation $ij$ as $b_{i,ij}$ then the link
$b_{i,ij} \rightarrow b_{i}$ is performed by}
\begin{equation}
    b_{i,ij} = (b_{i}-\beta)\frac{(1+z_{\mathrm{zeff},ij})^2}{(1+z_{\mathrm{zeff},i})^2} + \beta_i.
\end{equation}
The values for parameters $\alpha$ and $\beta$ are summarized in \cref{tab:bias_evolution} and measured from unblinded DESI DR1 data (see Appendix C of \cite{ChaussidonY1fnl}). 

\begin{table}
    \centering
    \caption{Values of the parameters used to describe the linear bias redshift evolution, measured from unblinded DESI DR1 data \cite{ChaussidonY1fnl}.}
    \begin{tabular}{ccc}
    \toprule
         &  $\alpha$& $\beta$\\
         \midrule
         LRG&  0.209& 1.415\\
         
         QSO&  0.237& 0.771\\
         
         ELG&  0.153& 1.541\\
         \bottomrule
    \end{tabular}
    \label{tab:bias_evolution}
\end{table}

\hl{Since the true inverse covariance matrix (precision matrix) of our observable is unknown, we estimate it from measurements performed on the mocks described in \cref{subsec:EZmocks}. This approximation leads to a biased estimate of the precision matrix, which we account for with the Hartlap factor \cite{Hartlap_factor_2006}. Another important correction is performed with the Percival factor \cite{Percival_factor_2014}, which accounts for the inflation of the errors on the estimated parameters due to the intrinsic noise of the estimated covariance matrix. Both corrections are used to rescale the precision matrix.}

Combining all samples with the current $k$-range and binning (linear binning) generates large data vectors. 
For example, the L+Q+LxQ+ExL+ExQ+E case has $n_p=13$ free parameters, $n=482$ data points, and $N_m=1000$ mocks were used to construct the covariance matrix, which yields Hartlap factor of $0.5$ and a Percival factor of 1.86. Therefore, we reduce the number of data points in the data vectors by using a `variable' binning scheme. We maintain a fine $k$-binning up to $k_\mathrm{mid}=2\times 10^{-2}\,\hmpc$ and use a coarse binning of $\Delta k=5\times 10^{-3}\,\hmpc$ up to $k_\mathrm{max}$. The variable binning scheme reduces the number of data points to $n=184$, resulting in a Hartlap factor $0.81$ and a Percival factor of 1.19.

\section{Results}
\label{sec:results}
In this section we discuss the expected gain from including the tracer cross-correlations when performing our analysis on the mocks and present the results from DESI DR1 data. We also discuss the DESI DR1 results in the context of the mocks.

\subsection{Expectation from the Mocks}
\label{subsec:mocks_results}
The constraints from the EZmocks are obtained in the same manner as we would for the data, with the exception of the multipoles being rescaled as described in \cref{subsec:EZmocks}. This test with the mocks is to assess the gain we expect from including the information of the cross-tracer samples in our analysis with the observed data. 

\cref{tab:mocks_summary} summarizes the results from fitting on the mean of 1000 EZmocks. For each case we present the central value for $\fnl$ obtained from likelihood profiling with \minuit\footnote{\url{https://github.com/scikit-hep/iminuit}} and their respective errors estimated from MCMC chains. Both are performed using the \desilike\footnote{\url{https://github.com/cosmodesi/desilike}} pipeline. In addition, we show the standard deviation of the MCMC chains for $\fnl$ as $\sigma(\fnl)$ which quantifies the gain from including the cross-correlations. The $\langle \chi^2 \rangle$ values in this table are obtained from performing profile fits on a subset of 100 EZmocks using a covariance matrix that excludes the mock used in the fitting.

\begin{table}
    \centering
    \caption{Results from fitting on the mean of 1000 EZmocks. The central values are obtained with \minuit, while the errors are estimated with MCMC. 
    Both are performed using the \desilike\, pipeline. In addition, the $\langle \chi^2 \rangle$ values are obtained from the mean of $\chi^2$ obtained from performing profile fits on a subset of 100 EZmocks using a covariance matrix that excludes the mock used in the fitting.}
    \begin{tabular}{lccccc}
\toprule
 Sample            &   $f^\mathrm{loc}_\mathrm{NL}$ &   $\sigma f^\mathrm{loc}_\mathrm{NL}$ &   $\langle \chi^2 \rangle$/dof \\
\midrule
 LRG               &  $-1.3_{-19.4}^{+23.4}$  &   21.86  &  20.8/22\\[0.65ex]

 QSO               &  $-1.4_{-9.6}^{+11.3}$   &   10.31  &  63.2/54\\[0.65ex]

 LRG+QSO           &  $-1.3_{-8.5}^{+10.1}$  &   9.31 &    85.0/77\\[0.65ex]

 L+Q+LxQ           &  $0.8_{-7.9}^{+9.4}$    &   8.62 &    103.8/102\\[0.65ex]

 L+Q+LxQ+ExL+ExQ+E &  $-0.8_{-7.4}^{+8.6}$   &   8.02 &    151.5/171\\
\bottomrule
\end{tabular}
    \label{tab:mocks_summary}
\end{table}

We now evaluate the gain we obtain from extending our sample to include additional clustering statistics. Our baseline is the result from the joint fit of the LRG+QSO sample, which yielded $\fnl=-1.3_{-8.5}^{+10.1}$ and $\sigma(\fnl)=9.31$. 
This precision is consistent with $\sigma(\fnl) \sim  9.0-9.1$ for the DESI DR1 data measured in \cite{ChaussidonY1fnl}. 
When we include the LRGxQSO sample in the measurement, we obtain $\fnl=0.8_{-7.9}^{+9.4}$ and $\sigma(\fnl)=8.62$,  an $\sim8\%$ gain. Furthermore, including the ELGxLRG, ELGxQSO, and ELG samples yields an additional 8\% gain, making the total a gain of $\sim16\%$ with $\fnl=-0.8_{-7.4}^{+8.6}$ and $\sigma(\fnl)=8.02$. The majority of the $\fnl$ information is contained within the QSO sample, which is further improved when combined with LRGs. We have demonstrated that a modest gain is possible by including all DESI tracers and their cross-correlations. 

\hl{The EZmocks are generated with $\fnl=0$, therefore we expect to recover an estimate of $\fnl=0$ within statistical fluctuations. However,} the mean $\fnl$ from our fits to the mocks show small offsets from the truth that are statistically significant considering 1000 mock realizations were used. The result in \cref{tab:mocks_summary} implies this shift of -1.3 for LRG+QSO is at $4.5 \sigma$ significance, where $\sigma$ is defined as $\sigma(\fnl)/\sqrt{1000}$, while LRG+QSO+LxQ and L+Q+LxQ+ExL+ExQ+E return $\pm 0.8$ at $\sim 3\sigma$. Therefore it appears that including the additional samples decreases this shift. These shifts are small compared to our uncertainty of the \hl{DR1 data ($\pm0.8/8.54 \approx0.1 \sigma$; using the tightest $\sigma(\fnl)$ from \cref{tab:data_summary}) and change sign depending on the exact set of tracers, therefore likely resulting in fluctuations rather than a net bias}. We therefore consider these small enough to ignore in the rest of our analysis, but as the precision of DESI $\fnl$ result improves, such shifts may indicate a need for additional systematic uncertainty.

\subsection{DR1 Results}
\label{subsec:results_data}
Now, we apply the same method on the DR1 data and present the $\fnl$ constraints from various data combinations. This is shown  in \cref{tab:data_summary}. The central and $\chi^2$ values are obtained with \minuit, while the errors are estimated from the MCMC chains. We continue using $\sigma(\fnl)$ to quantify the gain from including the cross-correlations in our analysis.

Our analysis yields $\fnl=-2.5_{-9.4}^{+9.5}$ and $\sigma(\fnl)=9.33$ for the LRG+QSO sample. Again, the result is well within $1\sigma$ of $\fnl=-3.6_{-9.1}^{+9.0}$ from \cite{ChaussidonY1fnl}. The small mismatch in central value and error bounds is expected since we include low redshift LRGs; in addition, we use $w_\mathrm{sys}$ when computing the effective redshift at which we evaluate the model.
When we include the LRGxQSO cross-power in the measurement, we obtain $\fnl=2.1_{-8.3}^{+8.8}$ and $\sigma(\fnl)=8.54$,  an $\sim9\%$ gain, very close to the expected 8\% gain from our mock tests. 

\hl{When including the ELGxLRG, ELGxQSO, and ELG measurements in addition to LRGxQSO, we obtain $\fnl=-4.9_{-9.6}^{+8.0}$ with $\sigma(\fnl)=8.64$, which is very similar to the precision when adding only the LRGxQSO cross-power. Therefore, we observe no further gains beyond including L+Q+LxQ for the DESI DR1. }

\hl{For the mocks, the expected gain of adding LRGxQSO was 8\%, and it reached a total of 16\% when including all measurements. We do not see this additional 8\% further gain in the data.  
From the mock tests done in \cref{subsec:data_in_context_of_mocks}, we find that 28\% of the mocks showed a smaller gain than the data for adding these extra ELG-related measurements, implying that the data result is explained by the statistical scatter.}

\hl{While there was no gain in precision, adding the extra measurements introduced a shift in the central $\fnl$ value of more than $0.5\sigma$. This shift is also allowed within the statistical fluctuations, which will be further discussed in the next section (\cref{subsec:data_in_context_of_mocks}).}

\begin{table}
    \centering
    \caption{Best-fit results on the DESI DR1 data. The central values are obtained with \minuit, while the errors are estimated with MCMC. Both are performed through the \desilike\, pipeline.}
   \begin{tabular}{llcc}
\toprule
 Sample            &   $f^\mathrm{loc}_\mathrm{NL}$ &   $\sigma f^\mathrm{loc}_\mathrm{NL}$ &  $\chi^2$/dof \\
\midrule
 LRG               &  $-3.9_{-22.0}^{+21.4}$ &  21.64  &   22.9/22 \\[0.65ex]

 QSO               &  $-2.1_{-10.1}^{+10.6}$   &  10.20  &   65.0/54  \\[0.65ex]
 
 LRG+QSO           &  $-2.5_{-9.4}^{+9.5}$    &   9.33  &   87.9/77 \\[0.65ex]

 L+Q+LxQ           &  $2.1_{-8.3}^{+8.8}$     &   8.54  &   113.6/102 \\[0.65ex]

 L+Q+LxQ+ExL+ExQ+E &  $-4.9_{-9.6}^{+8.0}$    &   8.64  &   195.6/171 \\
\bottomrule
\end{tabular}
    \label{tab:data_summary}
\end{table}

\begin{figure}
    \centering
    \includegraphics[width=\linewidth]{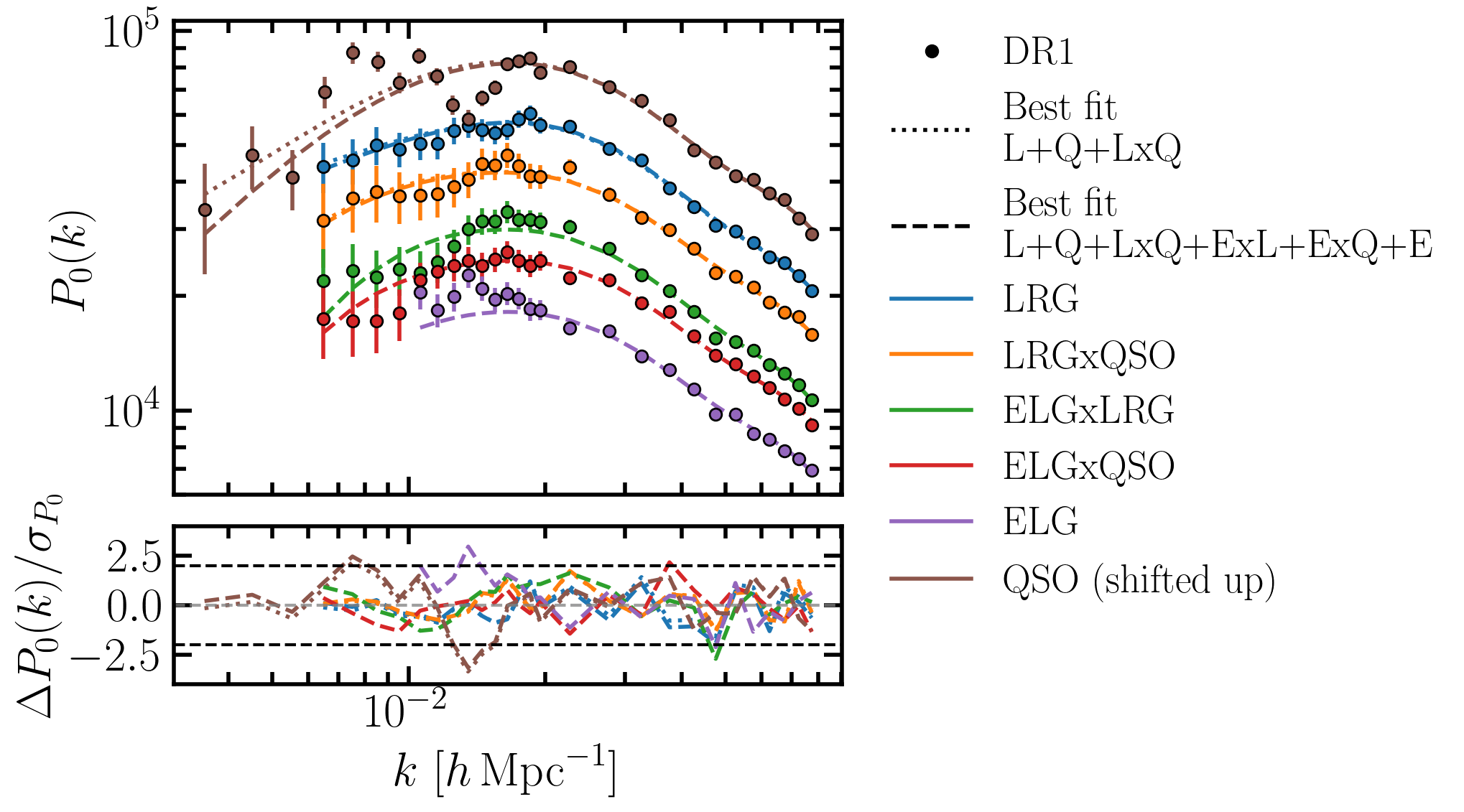}
    \caption{Monopoles for the best-fits from the L+Q+LxQ (dotted lines) and L+Q+LxQ+ExL+ExQ+E (dashed lines) samples compared to the observed power spectrum measurements from the DESI DR1 samples (circle markers). 
    The lower panel show the fractional errors from the difference between the best-fit models and the data.
    Note that the quadrupole of the auto-QSO is included in the analysis, but for convenience, it is not shown in this figure, since most of the constraining power is gained from the monopole. Also note that in the upper panel the QSO monopole was shifted vertically by a factor of two to avoid overlap with the LRGxQSO curve.}
    \label{fig:pk_data_bestfit}
\end{figure}

In \cref{fig:pk_data_bestfit} we show the power spectra for all the samples defined in \cref{subsec:dr1_data} and the best-fit models for the extended samples from \cref{tab:data_summary}. The errors in this figure correspond to the diagonal of the covariance matrix constructed from 1000 EZmocks as described in \cref{subsec:EZmocks}. In the lower panel we present the fractional errors from the difference between the best-fit models and the data. In both cases the best-fit models describe the data vectors within the error expected from the mocks.

\begin{figure}
    \centering
    \includegraphics[width=\linewidth]{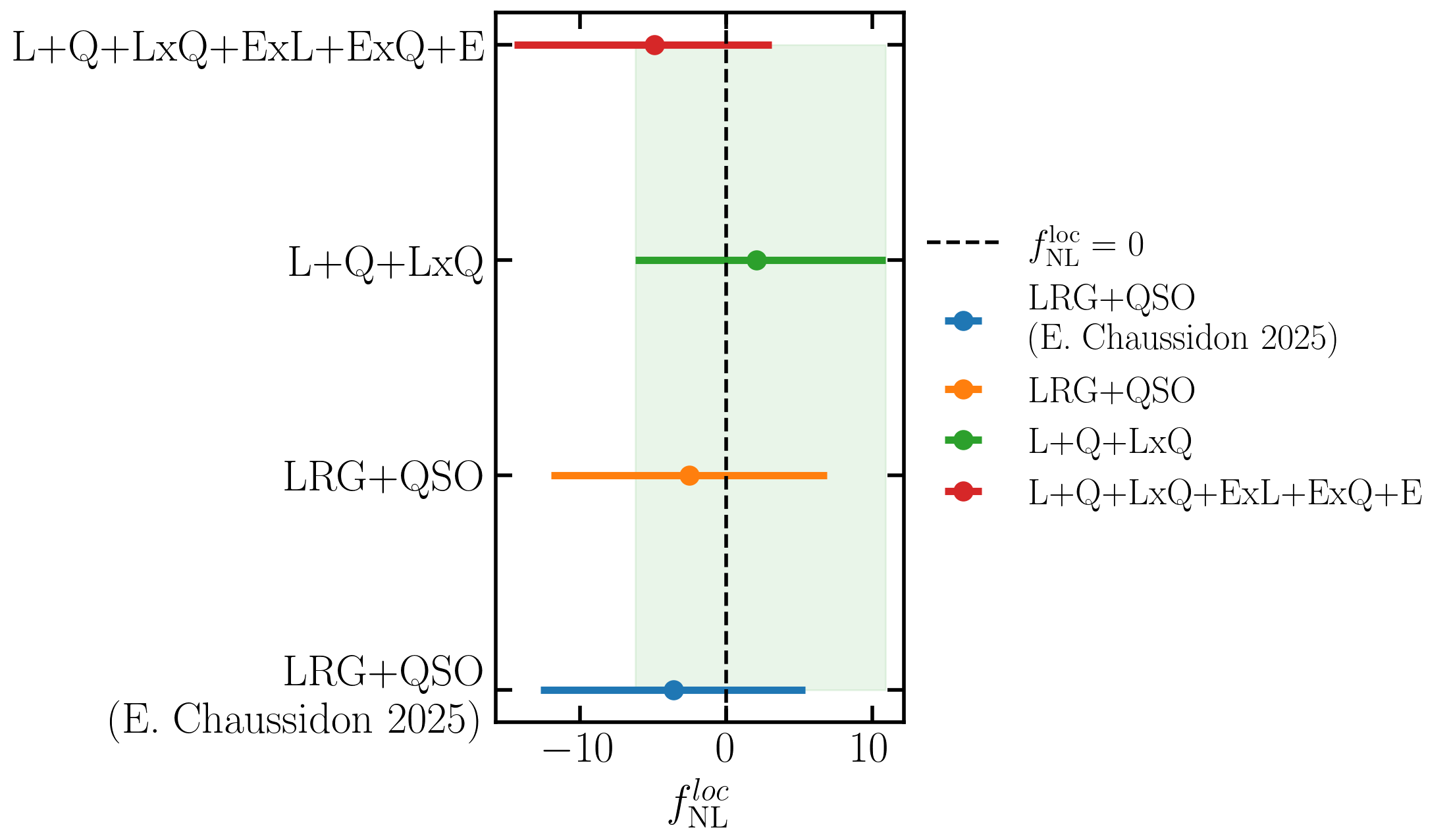}
    \caption{Constraints on $\fnl$ from including the cross-correlation of tracers compared to the joint constraint of auto-LRGs and QSOs. 
    The light green shaded region corresponds to the errors from the L+Q+LxQ measurement of $\fnl=2.1_{-8.3}^{+8.8}$ at 68\% confidence. 
    The LRG+QSO constraint of $\fnl=-2.5_{-9.4}^{+9.5}$ at 68\% confidence is represented by the orange marker, 
    while the L+Q+LxQ+ExL+ExQ+E constraint of $\fnl=-4.9_{-9.6}^{+8.0}$ at 68\% confidence is represented by the red marker.
    For comparison, we include the LRG+QSO constraint from \cite{ChaussidonY1fnl} of $\fnl=-3.6_{-9.1}^{+9.0}$ at 68\% confidence as a blue marker. The inclusion of the LRGxQSO cross-power yields an improvement in precision of $\sim9\%$.}
    \label{fig:whisker_data}
\end{figure}

We further summarize our key results from \cref{tab:data_summary} in \cref{fig:whisker_data}, which illustrates how the $\fnl$ constraints from including the cross-correlation of DESI tracers compare to the DESI DR1 combined analysis of LRGs and QSOs. 
In this figure we use our L+Q+LxQ constraint of $\fnl=2.1_{-8.3}^{+8.8}$ at 68\% confidence as the baseline and represent it with the green marker and its corresponding shaded region.
The L+Q+LxQ+ExL+ExQ+E constraint of $\fnl=-4.9_{-9.6}^{+8.0}$ at 68\% confidence is represented by the red marker, while our LRG+QSO measurement of $\fnl=-2.5_{-9.4}^{+9.5}$ at 68\% confidence is represented by the orange marker.
Also, for comparison, we include the LRG+QSO constraint from \cite{ChaussidonY1fnl} of $\fnl=-3.6_{-9.1}^{+9.0}$ at 68\% confidence as a blue marker.

\subsection{Contextualization of Results Based on Mocks}
\label{subsec:data_in_context_of_mocks}
The gain of $\sim8\%$ observed from the L+Q+LxQ+ExL+ExQ+E sample is less than the gain of $\sim16\%$ expected from the mocks. In this section, we test if this level of discrepancy is allowed within the statistical fluctuations. For this, we perform profile fits on 100 EZmocks using a covariance matrix that excludes the mock that is being fit. These are the same fits used to obtain the $\langle \chi^2 \rangle$ values reported in \cref{tab:mocks_summary}. 

\begin{figure}
    \centering
    \includegraphics[width=0.95\linewidth]{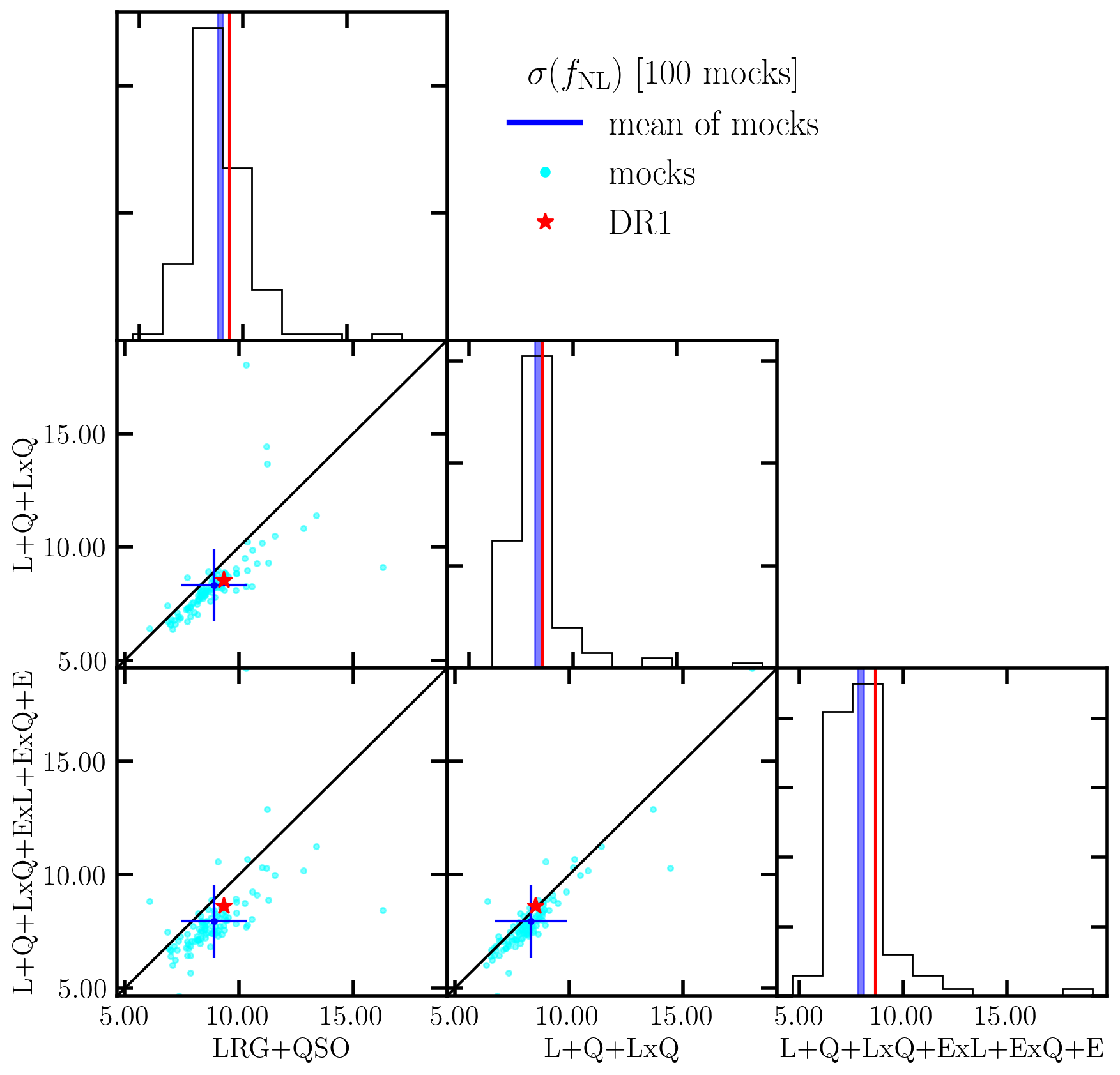}
    \caption{Comparison of the dispersion of the errors of $\fnl$,  $\sigma(\fnl)$, for the profiles fits performed on 100 EZmocks and DR1 data. Here, we show the best-fits from the mocks in cyan, the mean of the mocks in blue, and the data as a red star. The histograms on the diagonal show the distribution of $\sigma(\fnl)$, and the blue shaded region is the dispersion of the quantity $\sigma(\fnl)$ scaled by the square root of the number of mocks, while the red line is the data. The off-diagonal panels are plots that show the dispersion in $\sigma(\fnl)$ as recovered from the mocks, compared to the DR1 data. The error bars on the data points representing the mean of the mocks are the standard deviations for the $\sigma(\fnl)$ along each axis.}
    \label{fig:scatter_sigma_fnl}
\end{figure}

\cref{fig:scatter_sigma_fnl} compares $\sigma(\fnl)$ from the DR1 data to the dispersion from 100 EZmocks.
This figure compares the dispersion of the errors of $\fnl$,  $\sigma(\fnl)$, for the profiles fits performed on 100 EZmocks to the $\sigma(\fnl)$ of DR1 samples obtained with MCMC. The best-fits from the mocks are shown in cyan, the mean of the mocks in blue, and the data as a red star. The points with the error bars represent the mean and the standard deviations of the 100 mock $\sigma(\fnl)$ values along each axis.
The histograms on the diagonal show the distribution of $\sigma(\fnl)$, and the blue shaded region is the dispersion of the quantity $\sigma(\fnl)$ scaled by the square root of the number of mocks, while the red line is the data. The off-diagonal panels are plots that compare the precision $\sigma(\fnl)$ between different combinations of data sets. A precision gain in one combination relative to the other will show up as an offset from the diagonal lines. All of the DESI DR1 samples presented fall within the expected absolute as well as relative dispersion of $\sigma(\fnl)$ from the mocks.

\begin{figure}
    \centering
    \includegraphics[width=\linewidth]{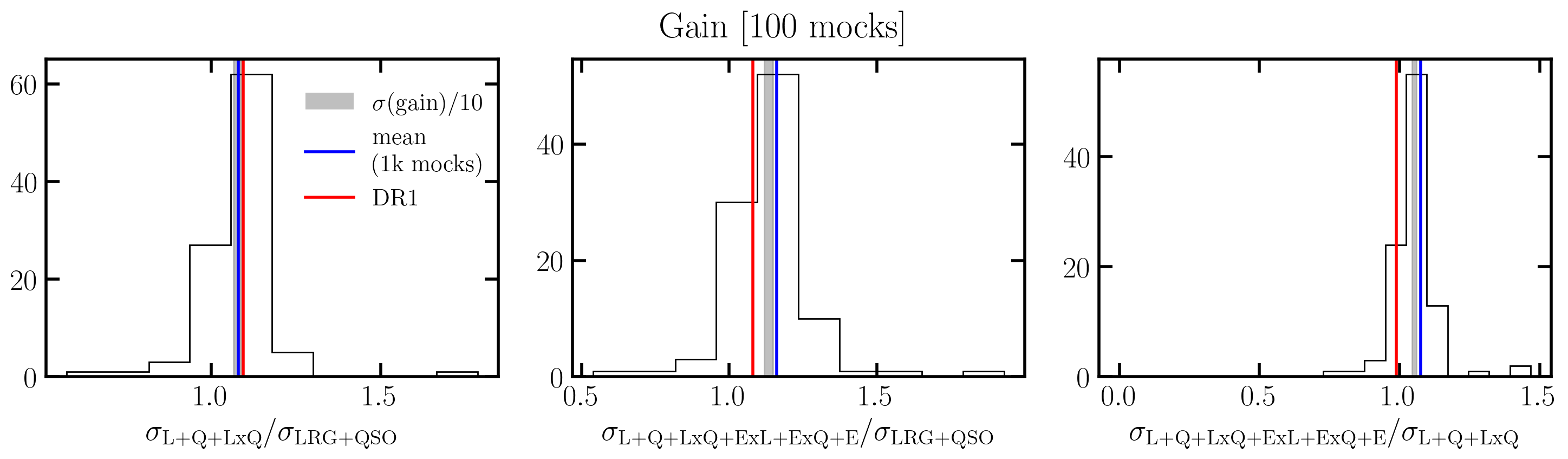}
    \caption{Histograms that show the expected gain in precision in the $\fnl$ measurement. The shaded regions represent the $1\sigma$ error of the mean of the 100 ratios of $\sigma(\fnl)$ around the mean of 100 best fits, $\sigma(\mathrm{gain}) / 10$. The blue vertical line represents the best-fit from fitting on the mean of 1000 EZmocks; while the red line represents the best-fit from the data.}
    \label{fig:gain_fnl}
\end{figure}

In \cref{fig:gain_fnl}, we rearrange the information in \cref{fig:scatter_sigma_fnl} to better illustrate the gain we observe for DR1 in comparison to the gains expected from the mocks.
The shaded regions represent the $1\sigma$ error of the mean of the 100 ratios of $\sigma(\fnl)$ around the mean of 100 best fits, $\sigma(\mathrm{gain}) / 10$.
Here, the blue line is from the best-fit from fitting on the mean of 1000 EZmocks. The red line represent the best-fit from the DESI DR1 data. 
\hl{The $\sim8\%$ gain observed in the data for L+Q+LxQ+ExL+ExQ+E is within the range of the mocks, with 28\% showing a smaller gain. Similarly, for the L+Q+LxQ sample, 71\% of the mocks show a smaller gain than observed in the data.}

\begin{figure}
    \centering
    \includegraphics[width=\linewidth]{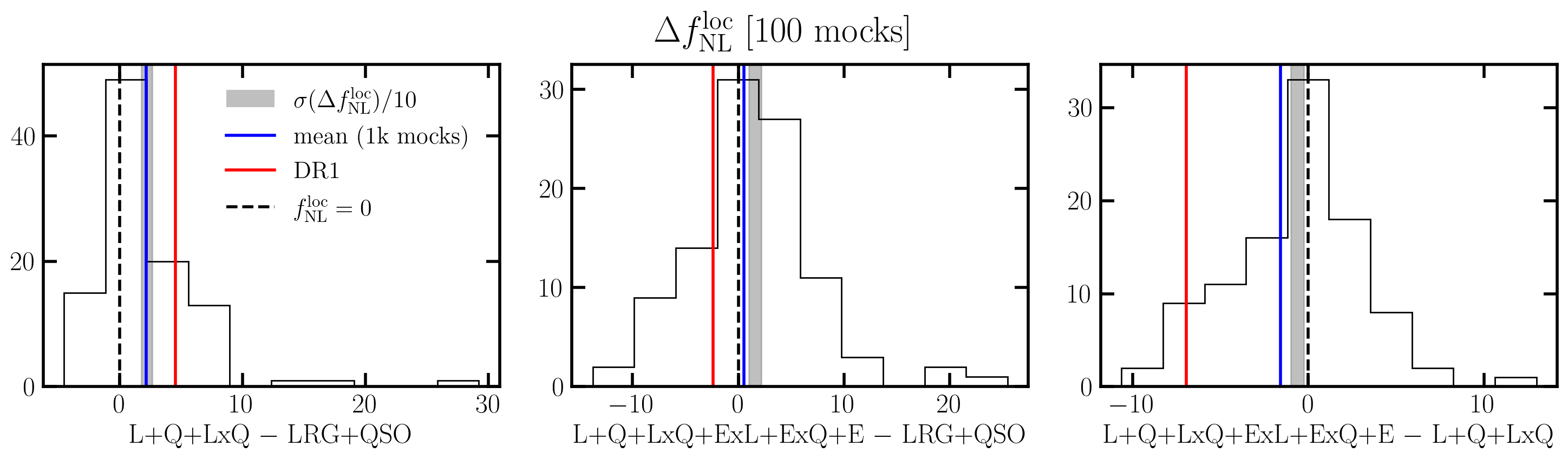}
    \caption{Histograms that show the difference between values of $\fnl$, $\Delta\fnl$, for different samples from the best-fits of 100 EZmocks. The shaded regions show the $1\sigma$ error of the mean of the 100 $\Delta\fnl$ around the mean of 100 best fits, $\sigma(\Delta\fnl) / 10$. The blue vertical line represents the best-fit from fitting on the mean of 1000 EZmocks; while the red line represents the best-fit from the data.}
    \label{fig:delta_fnl}
\end{figure}

In \cref{subsec:results_data}, from \cref{tab:data_summary} we found there is a shift of $\Delta\fnl=7$ in the central value of $\fnl$ observed in the L+Q+LxQ+ExL+ExQ+E sample. To understand the significance of this offset, we use the best-fits from 100 EZmocks and inspect the differences between the values of $\fnl$, $\Delta\fnl$, for different samples. The resulting histograms are shown in \cref{fig:delta_fnl}. Here, the shaded region shows the $1\sigma$ error of the mean of the 100 $\Delta\fnl$ around the mean of 100 best fits, $\sigma(\Delta\fnl) / 10$. The blue vertical line represents the best-fit from fitting on the mean of 1000 EZmocks. The blue line is within \hl{$2\sigma$ of} the mean of the best fits (the center of the gray shade), as expected. The offset of the gray shade from the expected truth at $\fnl=0$ shows any bias we may detect from the mock data.
\hl{We find that 6\% (66\%) of the mocks produce a more extreme $\Delta\fnl$ shift than observed in the data when comparing the L+Q+LxQ+ExL+ExQ+E sample to the L+Q+LxQ (LRG+QSO) sample.}

The large number of mocks and the variance cancellations between each pair of samples allows us detect a relative difference in $\fnl$ at a high significance; we find $\Delta\fnl = 2.1$, between the L+Q+LxQ  and LRG+QSO mock measurements at $4.5\sigma$ significance, although the underlying $\fnl$ for the mocks are all $\fnl=0$.
In detail, one can see in \cref{tab:mocks_summary} that the mean $\fnl$ measurement for L+Q+LxQ is slightly less biased than LRG+QSO, i.e. slightly more robust against the bias.
This may imply a potential relative bias of 2.1 existing between L+Q+LxQ  and L+Q of the data measurements as well.
Even if this shift is only specific to the mocks and did not propagate to the data measurements, if we were to subtract $\Delta\fnl = 2.1$ from the histogram in the left hand panel of \cref{fig:delta_fnl}, the shift measured on the data (red line) would be consistent with the statistical fluctuations observed in the EZmocks.
We therefore conclude that our DR1 $\fnl$ best-fits for LRG+QSO and L+Q+LxQ differ by an amount well within the expectation from statistical scatter.

\subsection{Comparisons to recent $\fnl$ measurements}
\label{subsec:results_comparisons}
Since the public release of DR1 \cite{desicollaboration2025datarelease1dark} and the publication of \cite{ChaussidonY1fnl} there have been $\fnl$ measurements by \cite{chudaykin2025reanalyzingdesidr13} and \cite{fondi2026assemblybiaslocalprimordial}. 
We compare our results with theirs using the same simulation-based values they adopted for the assembly bias parameter $p$.
In \cite{chudaykin2025reanalyzingdesidr13} they used the power spectrum and bispectrum to constrain $\fnl=-0.1^{+7.4}_{-7.4}$ using $p=0.55$ (preferred by stellar-mass selected samples as described in \cite{Barreira_2020a,Barreira_2020b}) for the LRGs and kept $p=1.6$ for the QSOs. In \cite{fondi2026assemblybiaslocalprimordial} they found a physically motivated value of $p=1.4$ for DESI QSOs, which led to a tighter constraint from only QSOs of $\fnl=-3.3^{+9.2}_{-9.2}$ (a $\sim12\%$ gain when compared to $\fnl=-2^{+11}_{-10}$ from \cite{ChaussidonY1fnl} with $p=1.6$).

Using the settings layed out in this paper, with the L+Q+LxQ sample we found $\fnl=2.1_{-8.3}^{+8.8}$. When we use the updated values of $p=0.55$ \cite{Barreira_2020a, Barreira_2020b} and $p=1.4$ \cite{fondi2026assemblybiaslocalprimordial} we obtain $\fnl=1.9^{+7.6}_{-6.8}$,\footnote{As in \cite{fondi2026assemblybiaslocalprimordial} we do not recompute the OQE weights with updated values of $p$. This is computationally expensive as it would require recomputing power spectra, window matrix and integral constraints.} which is the best constraint to date (from power spectra alone). However, we note that the value of $b_\phi$ for the LRGs is not well established and will lead to upcoming works \cite{miguel_icaza_lizaola}. 

To fairly compare our results to \cite{chudaykin2025reanalyzingdesidr13} we update our constraint to use $p=0.55$ for LRGs and keep $p=1.6$ for the QSOs, which yields $\fnl=-2.4^{+9.0}_{-8.3}$ for LRG+QSO and $\fnl=2.0^{+8.2}_{-7.7}$ for L+Q+LxQ. Between these results there is a $\sim7\%$ gain in precision. This is comparable to the $\sim 15\%$ gain they obtain when including the bispectrum and all other DESI tracers (BGS, LRGs, high-z ELGs) into their measurement.
We recognize that the gain observed in both cases might not be independent; however, we would like to highlight that the inclusion of LRGxQSO is robust and it is as significant as including higher order statistics.

\section{Conclusion}
\label{sec:conclusion}
In this work, we investigated the inclusion of the cross-correlation between DESI DR1 LRGs and QSOs in the PNG analysis. We also include DESI DR1 ELGs through their cross-correlation with other DESI DR1 tracers.
We quantified the gain from including these cross-correlations and validated them with EZmocks that match the large scale galaxy clustering of the DESI DR1 samples we use to constrain $\fnl$. 

\begin{itemize}
    \item DESI DR1 ELGs show significant excess clustering at large scales even after “cleaning” the imaging systematics \cite{Rosado-Marín_2025,KP3s15-Ross}. We believe that these are residual systematics to be understood. Hence, to avoid residual systematics biasing our constraints, we make a conservative choice, at the cost of precision, to cut the power spectrum of ELGs to $k_\mathrm{min}=0.01\,\hmpc$.
    
    \item Using the constraint from the LRG+QSO joint sample, $\fnl=-2.5_{-9.4}^{+9.5}$ at 68\% confidence, as a baseline we quantify the gain we obtain for each sample combination: 
    \begin{itemize}
        \item The gain from including the LRGxQSO sample is $\sim9\%$ for the DR1 data which yields $\fnl=2.1_{-8.3}^{+8.8}$ at 68\% confidence, consistent with the $\sim8\%$ gain expected from the EZmocks.
        \item  The expected gain from including the ELGs and other cross-correlations is $\sim16\%$, compared to the baseline, as quantified by the analysis performed on the EZmocks. On the other hand, the gain is of $\sim8\%$ for the DR1 data, with $\fnl=-4.9_{-9.6}^{+8.0}$ at 68\% confidence, \hl{which is almost the same as the gain from adding only LRGxQSO}. The smaller gain observed in the data is well within the dispersion of the 100 $\fnl$ measurements we performed on the mocks.
    \end{itemize}

    \item Including the cross-correlation between LRGs and QSOs can robustly improve the DESI DR1 $\fnl$ constraint. 
    Therefore, we recommend that all future DESI PNG studies include this cross-correlation in their analysis. The upcoming DESI DR2 PNG analysis will adopt this strategy by default.

    \item \hl{While no gain is obtained with the inclusion of ELGs and their cross-correlations with LRGs and QSOs for DR1 is consistent with the statistical scatter, we recommend future DESI studies revisit, as a more complete ELG sample may provide a gain comparable to that expected from the mocks.} 
    
\end{itemize}

This work will be extended to the DESI DR2 PNG analysis. There, we will include the covariance of the QSO sample with the rest of the tracers. Also, we will project the expected gain from performing a similar PNG analysis on the Y5 footprint. In addition, more rigorous imaging systematics tests will be conducted to optimally mitigate residual imaging systematics for PNG analysis, at large scales.

\section*{Acknowledgements}
AJRM and H-JS acknowledge support from the U.S. Department of Energy, Office of Science, Office of High Energy Physics under grant No. DE-SC0023241.

This material is based upon work supported by the U.S. Department of Energy (DOE), Office of Science, Office of High-Energy Physics, under Contract No. DE–AC02–05CH11231, and by the National Energy Research Scientific Computing Center, a DOE Office of Science User Facility under the same contract. Additional support for DESI was provided by the U.S. National Science Foundation (NSF), Division of Astronomical Sciences under Contract No. AST-0950945 to the NSF’s National Optical-Infrared Astronomy Research Laboratory; the Science and Technology Facilities Council of the United Kingdom; the Gordon and Betty Moore Foundation; the Heising-Simons Foundation; the French Alternative Energies and Atomic Energy Commission (CEA); the National Council of Humanities, Science and Technology of Mexico (CONAHCYT); the Ministry of Science, Innovation and Universities of Spain (MICIU/AEI/10.13039/501100011033), and by the DESI Member Institutions: \url{https://www.desi.lbl.gov/collaborating-institutions}.

The DESI Legacy Imaging Surveys consist of three individual and complementary projects: the Dark Energy Camera Legacy Survey (DECaLS), the Beijing-Arizona Sky Survey (BASS), and the Mayall z-band Legacy Survey (MzLS). DECaLS, BASS and MzLS together include data obtained, respectively, at the Blanco telescope, Cerro Tololo Inter-American Observatory, NSF’s NOIRLab; the Bok telescope, Steward Observatory, University of Arizona; and the Mayall telescope, Kitt Peak National Observatory, NOIRLab. NOIRLab is operated by the Association of Universities for Research in Astronomy (AURA) under a cooperative agreement with the National Science Foundation. Pipeline processing and analyses of the data were supported by NOIRLab and the Lawrence Berkeley National Laboratory. Legacy Surveys also uses data products from the Near-Earth Object Wide-field Infrared Survey Explorer (NEOWISE), a project of the Jet Propulsion Laboratory/California Institute of Technology, funded by the National Aeronautics and Space Administration. Legacy Surveys was supported by: the Director, Office of Science, Office of High Energy Physics of the U.S. Department of Energy; the National Energy Research Scientific Computing Center, a DOE Office of Science User Facility; the U.S. National Science Foundation, Division of Astronomical Sciences; the National Astronomical Observatories of China, the Chinese Academy of Sciences and the Chinese National Natural Science Foundation. LBNL is managed by the Regents of the University of California under contract to the U.S. Department of Energy. The complete acknowledgments can be found at \url{https://www.legacysurvey.org/}.

Any opinions, findings, and conclusions or recommendations expressed in this material are those of the author(s) and do not necessarily reflect the views of the U. S. National Science Foundation, the U. S. Department of Energy, or any of the listed funding agencies.

The authors are honored to be permitted to conduct scientific research on Iolkam Du’ag (Kitt Peak), a mountain with particular significance to the Tohono O’odham Nation.

\section*{Data Availability}
\label{sec:dataavail}

The data will be uploaded to Zenodo before publication\footnote{\url{https://zenodo.org/uploads/}}.

% references
\bibliographystyle{JHEP}
\bibliography{refs,DESI2024_updated15Aug, must} 

\clearpage
\appendix

\section{PNG constraints from ELGs}
As discussed in \cref{subsec:parameter_estimation}, DESI DR1 ELGs contain large-scale residual systematics. Including scales larger than $k=1\times10^{-2}\,\hmpc$ in our analysis would have rendered our analysis unreliable and positively biased our $\fnl$ constraint. Thus, we defined a conservative $k$-range for the ELG data vector. This decision came with a cost in precision to a sample that already has poor constraining power. For this reason, we did not include any results from ELGs only into the main text and present them here.

In \cref{tab:mocks_elgs_summary,tab:data_elgs_summary} we include the best-fits from the ELG sample and the L+Q+LxQ+E sample. In \cref{tab:mocks_elgs_summary} we 
show the best-fit results from fitting on the mean of 1000 EZmocks, same as \cref{tab:mocks_summary}. On the other hand, \cref{tab:data_elgs_summary} shows the best-fit results for the DESI DR1 data, similar to \cref{tab:data_summary}. As expected, the constraining power from ELGs only is poor. In addition, the best-fit is a poor fit to the data as shown by the $\chi^2$ values.

\begin{table}
    \centering
    \caption{This table follows \cref{tab:mocks_summary}, but includes results for the ELG and L+Q+LxQ+E samples.}
    \begin{tabular}{lccccc}
\toprule
 Sample            &   $f^\mathrm{loc}_\mathrm{NL}$ &   $\sigma f^\mathrm{loc}_\mathrm{NL}$ &   $\langle \chi^2 \rangle$/dof \\
\midrule
 ELG               &  $9.9_{-145.0}^{+66.0}$  &   103.63 &  14.1/18\\[0.65ex]

 LRG+QSO           &  $-1.3_{-8.5}^{+10.1}$  &   9.31 &    85.0/77\\[0.65ex]

 L+Q+LxQ           &  $0.8_{-7.9}^{+9.4}$    &   8.62 &    103.8/102\\[0.65ex]

 L+Q+LxQ+E         &  $0.9_{-7.6}^{+9.3}$    &   8.40 &    116.9/121\\[0.65ex]

 L+Q+LxQ+ExL+ExQ+E &  $-0.8_{-7.4}^{+8.6}$   &   8.02 &    151.5/171\\
\bottomrule
\end{tabular}
    \label{tab:mocks_elgs_summary}
\end{table}

\begin{table}
    \centering
    \caption{This table follows \cref{tab:data_summary}, but includes results for the ELG and L+Q+LxQ+E samples.}
   \begin{tabular}{llcc}
\toprule
 Sample            &   $f^\mathrm{loc}_\mathrm{NL}$ &   $\sigma f^\mathrm{loc}_\mathrm{NL}$ &  $\chi^2$/dof \\
\midrule
 
 ELG               &  $-41.7_{-32.9}^{+26.5}$ &   29.83 &   33.0/18 \\[0.65ex]

 LRG+QSO           &  $-2.5_{-9.4}^{+9.5}$    &   9.33  &   87.9/77 \\[0.65ex]

 L+Q+LxQ           &  $2.1_{-8.3}^{+8.8}$     &   8.54  &   113.6/102 \\[0.65ex]

 L+Q+LxQ+E         &  $-2.8_{-8.2}^{+8.8}$    &   8.43  &   150.4/121 \\[0.65ex]

 L+Q+LxQ+ExL+ExQ+E &  $-4.9_{-9.6}^{+8.0}$    &   8.64  &   195.6/171 \\
\bottomrule
\end{tabular}
    \label{tab:data_elgs_summary}
\end{table}

%\input{affiliations}
% Don't change these lines
%\label{lastpage}
\end{document}